\documentstyle[12pt]{article}
\newcommand{\thickone}{\mbox{$1\!\!1$}}
\newcount\subno      
\newcount\subsubno   
\newcount\appno      
\newcount\appeqno    
\newcount\tableno    
\newcount\figureno   
\def\nsection#1
   {
  \vskip0pt plus.1\vsize\penalty-250
    \vskip0pt plus-.1\vsize\vskip24pt plus12pt minus6pt
    \subno=0 \subsubno=0
    \addtocounter{section}{1}
    \noindent {\bf \thesection. #1\par}
    \noindent}
\def\nsubsecnn#1
   {\vskip-\lastskip
    \vskip24pt plus12pt minus6pt
    \noindent {\bf #1\par}
    \medskip
    \noindent}
\def\subsection#1
   {\vskip-\lastskip
    \vskip24pt plus12pt minus6pt
    \bigbreak
    \global\advance\subno by 1
    \subsubno=0
    \noindent {\sl \thesection. \the\subno. #1\par}
    \nobreak
    \medskip
    \noindent}
\def\appendix#1
   {\vskip0pt plus.1\vsize\penalty-250
    \vskip0pt plus-.1\vsize\vskip24pt plus12pt minus6pt
    \subno=0
    \global\advance\appno by 1
    \noindent {\bf Appendix  #1\par}
    \bigskip
    \noindent}
\begin{document}
\setcounter{page}{1}
\pagestyle{plain}
\setcounter{equation}{0}
%
%
%
\ \\[4mm]
\begin{center}
    {\bf EXACT STEADY STATES OF DISORDERED} \\[2mm]
  {\bf HOPPING PARTICLE MODELS } \\[2mm]
 {\bf WITH PARALLEL AND ORDERED SEQUENTIAL DYNAMICS}\\[12mm]
\end{center}
\begin{center}
\normalsize
M.\ R.\ Evans\footnote{
Royal Society University Research Fellow}\\[8mm]
{\it Department of Physics and Astronomy\\
        University of Edinburgh\\ Mayfield Road, Edinburgh EH9 3JZ,
U.K.}
\end{center}
\ \\[8mm]
\noindent {\bf Abstract:} A
one-dimensional driven lattice gas with disorder in the particle
hopping probabilities is considered.
 It has previously been shown that in the version
of the model with random sequential updating, a phase transition
occurs from a low density inhomogeneous phase to a high density
congested phase. Here the steady states for both parallel (fully
synchronous) updating and ordered sequential updating are solved
exactly and the phase transition shown to persist in both cases.
For parallel dynamics and forward ordered sequential dynamics 
the phase transition occurs at the same density
but  for backward ordered sequential dynamics it occurs
at a higher density.
In both cases the critical density is higher than that for
random sequential dynamics.
  In all the models studied the steady state
velocity is related to the fugacity of a Bose system
suggesting a principle of minimisation of velocity.
A generalisation of the dynamics where the hopping
probabilities depend on the number of empty sites in front of the
particles, is also solved exactly in the case of 
parallel updating.
 The models have  natural interpretations
as  simplistic descriptions of traffic flow. The relation to more
sophisticated traffic flow models is discussed.
\\[6mm]
\noindent Date: 29/4/1997 
\medskip 
\noindent Submitted to Journal of Physics A.
\\[3mm]
\begin{flushleft}
\parbox[t]{3.5cm}{\bf Key words:} asymmetric  exclusion process,
random rates,  steady state, parallel dynamics, phase transition,
Bose condensation, traffic flow
\parbox[t]{12.5cm}{ }
\\[2mm]
\parbox[t]{3.5cm}{\bf PACS numbers:}    05.40+j, 05.60+w, 64.60Cn, 89.40+k\\[2mm]
\parbox[t]{3.5cm}{\bf Short Title:} Disordered Hopping Particle Models 

\end{flushleft}
\normalsize
\thispagestyle{empty}
\mbox{}
\pagestyle{plain}
%
%
%
\newpage
\baselineskip=18pt plus 3pt minus 2pt
\setcounter{page}{1}
\setcounter{equation}{0}
\nsection{Introduction} 
The asymmetric simple exclusion process (ASEP)
is an archetypal example of a driven diffusive system \cite{SZ,Spohn}
for which analytical results are possible, particularly in one
dimension \cite{DE}.  The model comprises particles which hop
stochastically in a preferred direction with hard core exclusion
imposed. The model has a natural interpretation as a simplistic
description of traffic flow on a one lane road and indeed forms the
basis for more sophisticated traffic flow models \cite{BML,Nagel}. In
particular one may cite variations of the model proposed originally by
Nagel and Schreckenberg \cite{NS,MWR,SSNI,Naga1,Naga2, NP}.

However, a basic difference between the original ASEP and traffic flow
models lies in the updating scheme. In the mathematical literature the
ASEP is usually defined in continuous time or, equivalently for the
purposes of simulation, by a random sequential updating scheme where
for each update a particle is selected at random.  In contrast, when
simulating traffic flow parallel updating is usually employed for
reasons both idealistic--- parallel dynamics provides a perhaps more
faithful representation of real traffic--- and pragmatic---parallel
dynamics yields economy of random numbers.

For random sequential dynamics a relative wealth of exact results on the ASEP
are now available \cite{Spohn,DE,GS,DEHP,SD,vLK,DEM,DEMall,Gunter2}, in
particular the steady states of various models have been constructed
using a matrix product ansatz \cite{DE,DEHP,DJLS,Sven,EFGM2,ER,DLS}.
This technique has been extended to a sublattice parallel updating
scheme \cite{Gunter,Haye,HS} and, in the case of open boundary
conditions, to an ordered sequential scheme \cite{RSS,HP}. However, for
fully parallel dynamics only a few exact results are known
\cite{SSNI,BPSS}.

In the present work we determine exactly the steady state of an ASEP
with disorder for both parallel dynamics and ordered sequential
dynamics, in the case of periodic boundary conditions. The disorder
takes the form of quenched random hopping probabilities assigned to
each particle.  Other types of disorder and inhomogeneities such as
random rates associated with lattice sites \cite{CV}, slow defect
sites \cite{JL,TZ,HS} or slow defect particles \cite{BerChine,Kirone}
have  been considered.  The random sequential version of the
disordered model considered here has previously 
been studied \cite{BFL,MRE,KF}
and a transition shown to occur between a low density inhomogeneous
phase, where a tailback forms behind the slowest particle, to a high
density congested phase.

  In the present work it will be demonstrated
that this transition persists under parallel and ordered sequential
updating. The difficulty in obtaining exact results for parallel dynamics and
ordered sequential dynamics lies in the construction of the transfer
matrix. Using a technique inspired by ref. \cite{HN} we explicitly
construct the transfer matrices in a convenient form that allows the
steady states to be demonstrated.

From the point of view of traffic flow the phase structure of the
disordered model is of interest since the disorder
 induces emergent jams at low
densities, whereas in other traffic flow models where a phase
transition occurs it is in the high density phase that the jams emerge
\cite{Naga1,NP,CK}.   The coarsening of the resulting jams has also been
studied \cite{BKR,Naga2,KF,KCW}. From a theoretical viewpoint it was
shown in the random sequential case that the transition has a strong
analogy with Bose condensation \cite{MRE} and that the
steady state velocity of a particle was equivalent to the
fugacity of an ideal Bose gas.  This analogy will be
pursued here for the parallel and ordered sequential cases
and it will be shown that the steady state velocity
remains related to the fugacity of a Bose system.

It is of interest to determine whether distinct updating schemes can
produce different behaviour. It turns out that the
the value of the critical
density may depend on the updating scheme.
It will be shown that the critical density is highest, implying the throughput most
efficient, for a backward ordered updating scheme where
the updating  sequence is opposite to the direction of flow of
particles.   In contrast, the random sequential updating scheme
yields the lowest critical density. 

Another key difference between  traffic models (e.g. the model
of \cite{NS}) and the
ASEP is that in traffic models particles may move a distance
 greater than one lattice
spacing, thus implying that the dynamics is not nearest neighbour in
the sense that cars are aware of cars several lattice sites ahead.  At
present analytical results are not generally available
 for the case of hopping more than one lattice site,
 although a step towards this goal has been made \cite{BPSS}. On
the other hand dynamics of range greater than nearest neighbour may be
mimicked by letting the probability with which a particle hops forward
depend on the number of empty lattice sites in front of it.
Indeed,  the concept of a braking
distance furnishes a natural interpretation for hopping probabilities
which increase as a function of the empty space in front.

In the present work we solve  the steady state of a generalisation
of the dynamics where the
disordered hopping probabilities are dependent on the empty space in
front, restricting our attention to parallel updating.

The paper is organised as follows. In section 2 the model and updating
schemes considered are defined and exact expressions for the steady
states presented. The proof of these expressions is somewhat technical
and is deferred until section 4.  Prior to that, in section 3, the
phase transition and the analogy with Bose condensation is analysed.  In
section 5 the generalisation of the model to the case of space
dependent hopping probabilities is considered and in section 6
conclusions are drawn.

\nsection{Model Definitions and Steady States}
In this work we study
asymmetric exclusion models where the particle hopping probabilities
are quenched random variables.  We consider $M$ particles labelled $\{
\mu =1,\ldots, M \}$ hopping on a one dimensional lattice of size $N$
sites labelled $\{ i =1,\ldots, N \}$ with periodic boundary
conditions (site $N+i$ = site $i$).
The random sequential version of the model
was considered in \cite{MRE}.
We now define three distinct variants of the model
according to the following
dynamics. \\ \\
\noindent {\bf Parallel Dynamics}: at each time-step
all particles $\{ \mu =1,\ldots , M \}$ simultaneously  attempt 
to hop forward  each with its own probability $p_\mu$.   A hop is only carried
out if the target site was empty {\em before} the update. 
Since no backwards hops
are permitted the question of what would
 happen if two particles simultaneously attempted
to hop onto the same site does not arise.\\
\\
\noindent {\bf Ordered Sequential Dynamics}: 
in this case a time-step
corresponds to updating each particle in a fixed sequence.  At each
update the relevant particle $\mu$ attempts a hop forward with
probability $p_\mu$.  We consider two sequences for the updating

\begin{description}
\item[   ]{\bf Forward updating} in which the order is 
$1, 2, \ldots M$.
\item[   ] {\bf Backward updating} in which the order is
$M, M-1, \ldots 1$
\end{description}

\noindent Since particles cannot overtake, the sequence of particles is 
preserved in all three variants.

\subsection{Expressions for the steady states}
In order to express the steady state  of each  variant
of the model defined above, we consider the
weight $F( n_1,n_2,\ldots, n_M)$ of a configuration
comprising 
 particle 1 followed by $n_1$ holes (empty lattice sites); particle 2 followed by $n_2$ holes
and so on.
The weights are related to probabilities
$P_N( \{n_{\mu}\})$ via a normalisation $Z_{N,M}$ 
defined by
\begin{equation}
P_N( \{n_{\mu}\}) =F( \{n_{\mu}\})/Z_{N,M}.
\label{normdef}
\end{equation}

We now present exact expressions (to be proven in later
sections) for the steady state of the three variants.
In each case the steady state weights have the factorised form
\begin{equation}
F( \{ n_\mu \}) = \prod_{\mu=1}^{M} f_{\mu} ( n_{\mu} )
\label{Ffac}
\end{equation}
The different variants have different expressions for
$f_\mu( n_\mu )$ as follows.\\
\noindent {\bf Parallel Dynamics}\\
In this case 
\begin{eqnarray}
f_\mu( n_\mu )
 &=& (1- p_\mu)\;\;\;\mbox{for}\;\;\; n_\mu=0 \nonumber \\
 &=& \left( \frac{1-p_\mu}{p_\mu} \right)^{n_\mu}
\;\;\;\mbox{for}\;\;\; n_\mu >0 
\end{eqnarray}
 which may be rewritten employing the usual Heaviside function
as
\begin{eqnarray}
f_\mu( n_\mu )
&=&(1-p_\mu)+ \left[ \left( \frac{1-p_\mu}{p_\mu} \right)^{n_\mu}
	-(1-p_\mu) \right] \theta(n_\mu) \label{fP} \; .
\end{eqnarray}
Note that the pure case, which has $p_\mu = p \; \forall \mu$
recovers a result  obtained in \cite{SSNI} through a combinatorial
argument.
\\
\noindent {\bf Forward Ordered Sequential Dynamics}\\ 
To define a steady state for ordered sequential dynamics one must
specify the point in the update sequence
to  which the steady state refers.  For
forward updating we define the steady state as that after the update
of particle $M$ (the final update of the sequence).
That is, the steady
state weights are those for finding the system in a
given configuration after the last update of a sequence, and before the
first update of the next sequence. In this case
\begin{eqnarray}
f_\mu( n_\mu ) &=& (1-p_\mu)+ \left[ \left( \frac{1-p_\mu}{p_\mu} \right)^{n_\mu}
	-(1-p_\mu) \right] \theta(n_\mu) \;\;\;\mbox{for}\;\;\; \mu \neq M
\nonumber \\
f_M( n_M ) &=&  \left( \frac{1-p_M}{p_M} \right)^{n_M}
\label{fF}
\end{eqnarray}
\\
\noindent {\bf Backward Ordered Sequential Dynamics}\\ 
In this case the steady state again refers to the weight for
finding the system in a given configuration after the final update of
the sequence, this time the update of particle $1$.

\begin{eqnarray}
f_\mu( n_\mu ) &=&  \left( \frac{1-p_\mu}{p_\mu} \right)^{n_\mu}
 \;\;\;\mbox{for}\;\;\; \mu \neq M
\nonumber \\
f_M( n_M ) &=& (1-p_M)+ \left[ \left( \frac{1-p_M}{p_M} \right)^{n_M}
	-(1-p_M) \right] \theta(n_M)
\label{fB}
\end{eqnarray}
Before proving the steady states (\ref{fP},\ref{fF},\ref{fB}) we
discuss the implications of the form (\ref{Ffac}).  \nsection{Phase
transition and analogy with Bose condensation}

A steady state of the factorised form (\ref{Ffac}) gives rise to the
possibility of a phase transition.  In \cite{MRE} the analogy with
Bose condensation was made. Let us review briefly the qualitative
aspects of the phase transition and the analogy.

 The two phases exhibited by the model are a congested phase that
exists at high density and an inhomogeneous phase that exists at low
density.  In the congested phase, the velocity of particles is limited
by the availability of empty
sites, whereas in the inhomogeneous phase
the velocity is limited by the hopping rate of the slowest
particle. Thus, the inhomogeneous phase may be pictured as comprising
two regions: a tailback behind the slowest particle (a high density
region) and empty space in front of the slowest particle (a low
density region). The analogy with Bose condensation is to think of the
empty sites as bosons and the state of a boson as determined by which
car it is immediately in front of.  Then in the congested phase the
bosons are thinly spread over all the Bose states (i.e. there are
uniformly small gaps between the particles) whereas in the inhomogeneous phase
the bosons are condensed in front of the slowest particle (there is a
large empty space in front of the slowest particle).

Before exploring the analogy with Bose condensation we shall
discuss the mathematics of how calculations are performed.  
We focus first on the case of parallel dynamics and extend
the results to ordered
sequential dynamics in section 3.3.

To calculate quantities of interest, in particular steady
state averages, one must first have at hand expressions for the
normalisation $Z_{N,M}$ of the weights (\ref{Ffac}) defined through
(\ref{normdef}). Due to the constraint of $N-M$ holes the
normalisation may be written
\begin{equation}
Z_{N,M} =
 \sum_{n_1,n_2 \ldots n_M}
\delta_{  \sum_{\mu} n_\mu,\ (N-M)}
 \prod_{\mu=1}^{M} f_\mu( n_\mu )
\label{norm}
\end{equation}
This quantity may be considered as the canonical partition function 
of a thermodynamic system and
in the standard way \cite{Huang} it may be written using an integral
representation of the delta function as
\begin{equation}
Z_{N,M} = \oint \, \frac{dz}{2\pi i} \ z^{-(N-M+1)}\ {\cal Z}(z)\; ,
\label{Zint}
\end{equation}
where the grand canonical partition function ${\cal Z}$ is given by
\begin{equation}
{\cal Z}(z) =
\prod_{\mu=1}^{M}
\left[ \sum_{n_\mu=0}^{\infty} z^{n_\mu} \ f_\mu(n_\mu) \right]\; .
\label{Zgc}
\end{equation}
For large $N,M$ (\ref{Zint}) is dominated by the saddle point of the
integral and $z^*$, the value of $z$ at the saddle point, is the
fugacity.

We now  calculate the velocity $v$, defined as the steady
 state average of the rate of hopping of a given particle.  Since
 the particles cannot overtake each other, the velocity is
 the same for any particle $\mu$. Taking $\mu = 1$ one finds
\begin{eqnarray}
v &=&
Z_{N,M}^{-1} \oint \, \frac{dz}{2\pi i} \ z^{-(N-M+1)}\
\left[ p_1 \sum_{n_1=1}^{\infty} z^{n_1} \ f_1(n_1) \right]
\prod_{\mu=2}^{M}\left[ \sum_{n_\mu=0}^{\infty} z^{n_\mu} \ f_\mu(n_\mu)
\right]
\label{vel1}\\
&\simeq&
\left[p_1 \sum_{n_1=1}^{\infty} (z^*)^{n_1} \ f_1(n_1) \right]
\left[ \sum_{n_1=0}^{\infty} (z^*)^{n_1} \ f_1(n_1) \right]^{-1}
\;\;\;\mbox{ for large}\;\;\; N,M 
\label{vel2} \\
&=& \frac{z^*}{1+z^*}\;,
\label{vel3}
\end{eqnarray}
where  the last equality results from
 performing the geometric series obtained
when (\ref{fP}) is inserted in (\ref{vel2}).

In the following the thermodynamic limit is defined by
\begin{equation}
N \rightarrow \infty \;\;\;\mbox{with}\;\;\; M =\rho N\; ,
\label{thermlim}
\end{equation}
with   the density $\rho$  held fixed.
In order to determine the fugacity one uses the condition
\begin{equation}
N-M = z \frac{\partial\ \mbox{ln}\, {\cal Z}}{\partial z}
\label{zgce}
\end{equation}
which is the saddle point condition for (\ref{Zint}) (or equivalently
 the condition that in the grand canonical ensemble the average number
 of empty sites is $N-M$).  Using (\ref{fP}) one finds that
 (\ref{Zgc}) becomes
\begin{equation}
{\cal Z}(z) =
\prod_{\mu=1}^{M}
\left[ 1-p_\mu + \frac{ (1-p_\mu)z}{p_\mu -(1-p_\mu)z} \right]\; .
\label{Zgc2}
\end{equation}
and (\ref{zgce}) yields
\begin{equation}
N-M = \frac{z}{1+z}\sum_{\mu=1}^{M} \frac{1}{p_\mu -(1-p_\mu)z}
\label{zgce2}
\end{equation}
Now using the relation between $z^*$ and $v$ (\ref{vel3}), one obtains
 in the thermodynamic limit
\begin{equation}
1-\rho = v(1-v) \frac{1}{N} \sum_{\mu=1}^{M}\ \frac{1}{p_\mu -v}
\label{vel4}
\end{equation}
\subsection{The disorder average}
In the following we shall consider the particle hopping probabilities
$p_\mu$ as quenched random variables drawn from a common distribution
\cite{KF}
\begin{equation}
{\cal P}(p) = \frac{ (\gamma +1)^{\ \ \ }}{(1-c)^{\gamma +1}}
	\, (p-c)^\gamma
\label{dist}
\end{equation}
with support on the interval 
$[ c , 1)$. Other distributions may also be considered
\cite{MRE}, but the qualitative behaviour is determined
by the power $\gamma$.

For the disordered system one wishes to obtain properties given
by a typical realisation of the disorder (here the particle hopping
probabilities), for example the typical velocity. Usually in the
theory of disordered systems one expects quantities such as $v$ to
self-average but quantities exponentially large in the system size,
such as $Z_{N,M}$, not to. That is, one expects as $N\rightarrow
\infty$, $v \rightarrow \overline{v}$ with probability one,
 where the bar indicates an average over the
particle hopping probabilities.  When working in the grand
canonical ensemble one has directly an equation for $v$
in the thermodynamic limit, 
(\ref{vel4}). Thus, with the assumption that $v$ is self-averaging,
one simply has to average (\ref{vel4}) to obtain an equation for the
typical velocity.

  Now, in the thermodynamic limit (\ref{thermlim}), the fraction of
particles with hopping probabilities between $p$ and $p+dp$ converges
to ${\cal P}(p) dp$.  Equally the velocity of the slowest particle
converges to $c$.  For the sake of clarity, however, it is convenient
to assume that the slowest particle, taken to be particle 1, has
velocity exactly equal to $c$.  Therefore (\ref{vel4}) may be replaced
by
\begin{equation}
1-\rho =  \rho (1-v) I(v) + \frac{\langle n_1 \rangle}{N}
\label{gce2}
\end{equation}
where
\begin{equation}
I(v)=\int_{c}^{1} dp \ {\cal P}(p) \frac{v}{p-v}
\label{int}
\end{equation}
Since the rhs of (\ref{zgce2}) is an increasing function of $z$ we
deduce that the rhs of (\ref{vel4}) is an increasing function of
$v$. However, $z$ was introduced in (\ref{Zint}) as a contour
integration variable therefore the saddle point $z^*$ must lie between
0 and any pole in the integrand of (\ref{Zint}).  Owing to this, the
maximum value $z$ can take is $c/(1-c)$ and the maximum value of $v$
is $c$.

If the integral (\ref{int}) diverges as $v \to c$ then (\ref{gce2})
can always be satisfied with $v < c$ and $< n_1 >/N$ zero. However, it
turns out that for the distribution (\ref{dist}) with $\gamma > 0$,
$I(v)$ is always finite as $v \rightarrow c$ from below. Therefore
when
\begin{equation}
I(c) < \frac{(1-\rho)}{\rho \ (1-c)}\; ,
\label{cond}
\end{equation}
which holds for $\rho$ below a critical value $\rho_c$,
 (\ref{gce2})
can only be satisfied with  $v \to c $ and $< n_1 >/N$ non-zero.
As (\ref{gce2}) expresses the constraint
in the number of holes, we see that in this 
case a finite fraction of the holes must
 reside in front of the slowest
particle.  Thus, we have the inhomogeneous low density phase.
On the other hand for $\rho > \rho_c$,
(\ref{gce2}) may be satisfied for $v < c$ and
$< n_1 >/N$ zero, in which case we have the high density congested phase.

  By this point the qualitative analogy with Bose condensation 
should be apparent.
However, we first remark that the exact mapping onto 
an ideal Bose gas found for
 random sequential dynamics \cite{vLK,MRE} no longer holds.
In the mapping
 $n_\mu$ is viewed as the occupation number of the $\mu^{\rm th}$ Bose
state; applying this to
Eq. \ref{fP} one sees that  in the parallel
update system an unoccupied Bose
state is penalised. This effective repulsion
between particles already occurs in the pure case
\cite{SSNI}. Secondly, for random sequential dynamics the velocity
was exactly equivalent to the fugacity of an ideal Bose gas
\cite{MRE}, whereas in the present case although $v$ is still an
increasing function of $z^*$, the relation is modified to
(\ref{vel3}).  Nevertheless, the analogy with Bose condensation
remains useful.  The slowest particle corresponds to the
Bose ground state and the distribution of particle hopping
probabilities ${\cal P}(p)$ corresponds to the density of states of
the Bose gas.  Thus the transition to the inhomogeneous phase
corresponds to a condensation of a finite fraction of bosons into the
Bose ground state (i.e. a finite fraction of the holes reside in
front of the slowest particle).

\subsection{Critical Behaviour}
In order to analyse the phase transition we require at our disposal
the asymptotic behaviour of (\ref{gce2}) as $v \to c$.  The expansion in
$\epsilon$ where
\begin{equation}
\epsilon = \ln\left[\frac{c}{v}\right]\simeq \frac{c-v}{c}
\;\;\;\mbox{as}\;\;\;v\to c
\end{equation}
 is carried out in appendix A. The result is
\begin{eqnarray}
\mbox{For}\;\;\; \gamma < 1
&I(v)& = \frac{(1+ \gamma)c}{\gamma (1-c)}
+(1+\gamma) \left( \frac{c}{1-c} \right)^{1+\gamma} \Gamma(1+\gamma) \Gamma(-\gamma)
\epsilon^\gamma + \ldots
\label{gamm<1} \\
\mbox{For}\;\;\; \gamma = 1
&I(v)& = \frac{2c}{(1-c)}+ 2\left( \frac{c}{1-c}\right)^2 
\epsilon \ln (\epsilon)  + \ldots
\label{gamm=1} \\
\mbox{For}\;\;\; \gamma > 1
&I(v)& = \frac{(1+ \gamma)c}{\gamma (1-c)}
- \frac{(1+ \gamma)c}{\gamma (1-c)}\left[ 1+ 
\frac{c}{(1-c)}\ \frac{\gamma}{(\gamma -1)}\right] \epsilon
+\ldots \label{gamm>1}
\end{eqnarray}
 we see from (\ref{gamm<1}--\ref{gamm>1}) that
$I(c)= (1+ \gamma)c/(\gamma (1-c))$
which implies, together with the condition for the transition (\ref{cond}),
that the critical value of $\rho$ is
\begin{equation}
\rho_c = \frac{\gamma}{\gamma +c+ \gamma c}\;.
\label{rhocP}
\end{equation}
One should note that for
that  $\gamma =0$ (a flat distribution of hopping probabilities)
 the transition
to a congested phase occurs at zero density.

As $\rho$ is increased above the critical value
the velocity decreases from $c$ according to
\begin{eqnarray}
\mbox{For}\;\;\; \gamma < 1 \;\;\; v &\simeq& c- \left[ \frac{ (1-c)^\gamma}{\rho_c^2\ c\ \Gamma(1+\gamma)
\ \mid \Gamma(-\gamma)\mid} \right]^{1/\gamma} (\rho-\rho_c)^{1/\gamma}+\ldots
\label{vexp1}
\\
\mbox{For}\;\;\; \gamma = 1 \;\;\;
 v &\simeq& c-  \frac{(1-c)}{2 \rho_c^2\ c} 
\frac{(\rho-\rho_c)}{\mid \ln (\rho-\rho_c)\mid } +\ldots
\label{vexp2}
\\
\mbox{For}\;\;\; \gamma > 1 \;\;\;
 v &\simeq& c-  \frac{(1-c)\gamma (\gamma-1)}{ \rho_c^2\ (1+\gamma)
\left[ c + (\gamma-1)(1-c) \right]} (\rho-\rho_c)  + \ldots
\label{vexp3}
\end{eqnarray}

One can also consider the current or throughput $J$ defined by
\begin{equation}
J=\rho\ v\;.
\end{equation}
In the inhomogeneous phase ($\rho < \rho_c$) the current increases
linearly with $\rho$ and $J=c \rho$.  Using the expansions
(\ref{vexp1}--\ref{vexp3}) the behaviour of the current in the high
density congested phase can be analysed.  Just above the transition
the current increases with $\rho$ if $\gamma < \gamma^*$ but decreases
if $\gamma > \gamma^*$ where
\begin{equation}
\gamma^*= \frac{1-c+c^2}{2(1-c)}
\left[\ 1+ \sqrt{ 1+4(1-c)c^2/(1-c+c^2)^2}\ \right]\;.
\end{equation}
Thus the maximum current is achieved in the
high density phase if $\gamma < \gamma^*$
and exactly at the phase transition if $\gamma > \gamma^*$.
However it does not appear that $\gamma^*$ is universal
because it depends on $c$ {\it i.e.} it depends on the particular
choice of ${\cal P}(p)$.

\subsection{Ordered Sequential Dynamics}

For forward sequential or backward sequential dynamics the velocity
 may be defined as the probability that the {\it first} particle in the
 updating sequence hops forward. It is easy to check that (\ref{vel3})
 is still the correct expression for the velocity in both these cases so
 that the velocity is always related to the fugacity in a simple way.
 By performing the sums in (\ref{Zgc}) one obtains for forward
 updating
\begin{equation}
{\cal Z}(z) =
\prod_{\mu=1}^{M-1}
\left[ 1-p_\mu + \frac{ (1-p_\mu)z}{p_\mu -(1-p_\mu)z} \right]
\left[ \frac{ p_M}{p_M -(1-p_M)z} \right]\; .
\label{Zgc2f}
\end{equation}
and for backward updating
\begin{equation}
{\cal Z}(z) =
\prod_{\mu=1}^{M-1}
\left[ \frac{ p_\mu}{p_\mu -(1-p_\mu)z} \right]\
\left[ 1-p_M + \frac{ (1-p_M)z}{p_M -(1-p_M)z} \right]\;.
\label{Zgc2b}
\end{equation}

For forward updating the saddle point value
of the fugacity $z$ and hence  the velocity will be the
same as
 as for parallel updating, implying the same
critical density.  However, for backward updating (\ref{Zgc2b})
gives a different saddle point value of $z$ to (\ref{Zgc2}),
 and it can be checked that (\ref{gce2}) is
modified to
\begin{equation}
\left( \frac{1-\rho}{\rho} \right)
= (1-v)\ I(v) - v\; ,
\end{equation}
 giving
\begin{equation}
\rho_c = \frac{\gamma}{c + \gamma}\; .
\label{rhocf}
\end{equation}
We see that for backward updating $\rho_c$ is greater
than  for parallel and forward updating
which share the same $\rho_c$ given by
 (\ref{rhocP}).
Since the velocity in the
low density phase is greater than that
in the high density phase, this means
that backward updating yields the 
best throughput. One might expect this because backward updating
increases the chance that the site in front of the next particle
to be updated has just been vacated.
One can also calculate the critical density for random sequential
dynamics \cite{MRE,KF} to find
$\rho_c = \gamma(1-c)/(c + \gamma)$ which is lower than both
 (\ref{rhocf}) and (\ref{rhocP}). This implies that random sequential
updating gives the poorest throughput.

\nsection{Proof of Steady State for Ordered Sequential and Parallel
Dynamics}
In all three variants of the model considered here the updating
rules comprises a determined  time-step. Therefore a
transfer matrix may be used to express the condition for the steady
state weights as
\begin{equation}
\sum_{{\cal C}'} T( {\cal C}, {\cal C}' ) F( {\cal C}')
= F( {\cal C})\; ,
\label{transfmat}
\end{equation}
where $T( {\cal C}, {\cal C}' )$ are the components of the
transfer matrix.
In (\ref{transfmat})
$T( {\cal C}, {\cal C}' )$ is the probability
of going from configuration ${\cal C}'$ to 
${\cal C}$ in one time-step and
$T( {\cal C}, {\cal C} )$ is the probability
of remaining in configuration ${\cal C}$ after
a time-step. A configuration
is specified by the hole occupation numbers
${\cal C} = \{n_1,\ldots, n_M \}$.
We see from (\ref{transfmat}) that
the steady state weights form an eigenvector  of the transfer matrix
with eigenvalue one.

In the construction of the transfer matrix that follows, it is
convenient to use an operator notation.  To each configuration $
\{n_1,\ldots, n_M \}$ is associated  a vector $ |\ n_1, \ldots ,
n_M \rangle $, to be referred to as the configuration
vector. The configuration vectors $ |\ n_1, \ldots , n_M \rangle $ form
an orthonormal basis for the vector space of configurations.  We write
the weight of a configuration as
\begin{equation}
F(  n_1, \ldots , n_\mu ) = \langle F |\  n_1, \ldots , n_\mu \rangle
\end{equation}
Thus, the the weight of a configuration $\{n_1,\ldots, n_M \}$
is the component in the
direction $\langle n_1, \ldots , n_M |$
of the bra vector $\langle F | $.

With this notation we may rewrite (\ref{transfmat}) as
\begin{equation}
\langle F | \hat{T} |\  n_1, \ldots , n_\mu \rangle
= \langle F  |\  n_1, \ldots , n_\mu \rangle
\label{TMop}
\end{equation}
where $\hat{T}$ is an operator acting on the space of configuration
vectors defined above: $\hat{T}$ acting on a given configuration
 vector  generates the  possible
configuration vectors before an update multiplied by the appropriate
transition probabilities.  We shall refer to $\hat{T}$ as the transfer
matrix as well as its components $T( {\cal C}, {\cal C}' )$.  Our task
now is to construct $\hat{T}$.

\subsection{Construction of the Transfer Matrices}
In this subsection we construct the transfer matrix for all three
variants of the dynamics. The desired form is of a trace of a product
of matrices, each of which contains as elements operators acting at
the relevant site \cite{HN}. The technique is most directly illustrated in the
case of ordered sequential dynamics, which we consider first in this
subsection.

The transfer matrix for a full time-step of ordered sequential
dynamics may be written as an ordered product of $M$ operators
$h_{\mu-1 \ \mu}$ corresponding to the update of each particle $\mu$
in sequence. Recalling that $\hat T$ acting on $|\ n_1, \ldots , n_\mu
\rangle $ generates the possible configurations leading to $|\ n_1,
\ldots , n_\mu \rangle$, one deduces that to correctly generate these
configurations we first act on $|\ n_1, \ldots , n_\mu \rangle$ with
an operator corresponding to the last update of the sequence, then
with an operator corresponding to the second last update and so on, so
that for the forward updating
\begin{equation}
\hat T_F =  \prod_{\mu=1}^{M} h_{\mu-1 \ \mu}\;.
\label{TF1}
\end{equation}
By similar reasoning,
 for  backward updating  the
transfer matrix $T_B$ is
\begin{equation}
\hat T_B = \prod_{\mu=1}^{M} h_{M-\mu \ M-\mu+1}\;.
\label{TB1}
\end{equation}
In both cases the operator $h_{\mu-1 \ \mu}$ is given by
\begin{equation}
h_{\mu-1 \ \mu} = \thickone - p_\mu a_\mu a_\mu^\dag + p_\mu
a_{\mu-1}^{\dag} a_\mu
\label{hmu}
\end{equation}
where $a_\mu$ is a raising operator and
$a_\mu^{\dag}$ is a lowering operator acting
on the vector space spanned by 
 $ | n_{1}, \ldots ,n_{M} > $:
\begin{eqnarray}
a_\mu  | n_{1}, \ldots, n_\mu, \ldots, n_{M} >
&=& | n_{1}, \ldots, n_\mu +1, \ldots, n_{M} > 
\label{raise} \\
a_\mu^{\dag}  | n_{1}, \ldots, n_\mu  , \ldots, n_{M} >
&=& | n_{1}, \ldots, n_\mu -1, \ldots, n_{M} > \;\;\;\mbox{if}\;\;\
n_\mu > 0  \label{lower}\\
&=& 0 \;\;\;\mbox{if}\;\;\
n_\mu = 0 \; ,
\label{lower2}
\end{eqnarray}
and one may verify that
$$
h_{\mu-1 \ \mu} | n_{1}, \ldots ,n_{M} > = (1 - p_\mu \theta(n_\mu)) |
n_{1}, \ldots ,n_{M} > + p_\mu \theta( n_{\mu-1})| n_{1}, \ldots
,n_{\mu-1} -1, n_\mu + 1, \ldots, n_{M} > \; ,
$$
correctly giving the configurations leading to $| n_{1}, \ldots ,n_{M}
>$ after the update of particle $\mu$.

To proceed further we employ techniques used in the study of
integrable models \cite{HN}. First it is easy to check that
(\ref{hmu}) may be rewritten in two ways as
\begin{eqnarray}
h_{\mu-1\ \mu} &=&
\begin{array}{c}
 ( \thickone,\ a^\dag_{\mu-1}) \\
\mbox{ }
\end{array}
\left( \begin{array}{c}
	\thickone -p_\mu a_\mu a^\dag_\mu \\
	p_\mu a_\mu
	\end{array} \right) 
\label{hmu1}\\
&=&
\begin{array}{c}
 ( \thickone -p_\mu a_\mu a^\dag_\mu,\ 	p_\mu a_\mu) \\
\mbox{ }
\end{array}
\left( \begin{array}{c}
	 \thickone\\
	 a^\dag_{\mu-1}
	\end{array} \right) 
\label{hmu2}
\end{eqnarray}
Equations (\ref{hmu1}) and (\ref{hmu2}) are to be read as a scalar
product in an auxiliary space (see below).  On inserting (\ref{hmu1})
into (\ref{TF1}), $\hat T_F$ becomes
\begin{equation}
\hat T_F =
\begin{array}{c}
 ( \thickone,\ a^\dag_{M}) \\
\mbox{ }
\end{array}
\left[
\prod_{\mu=1}^{M-1}
\left(
\begin{array}{cc}
\thickone -p_\mu a_\mu a^\dag_\mu & 
 a_\mu^\dag-p_\mu a_\mu a^\dag_\mu a^\dag_\mu \\
p_\mu a_\mu & p_\mu a_\mu a_\mu^\dag
\end{array}
\right)
\right]
\left( \begin{array}{c}
	\thickone -p_M a_M a^\dag_M \\
	p_M a_M
	\end{array} \right) 
\label{TF2}
\end{equation}
which may be rewritten as
\begin{equation}
\hat T_F = \mbox{Trace} \left[
\prod_{\mu=1}^{M-1}
\left(
\begin{array}{cc}
\thickone -p_\mu a_\mu a^\dag_\mu & 
 a_\mu^\dag-p_\mu a_\mu a^\dag_\mu a^\dag_\mu \\
p_\mu a_\mu & p_\mu a_\mu a_\mu^\dag
\end{array}
\right)
\left(
\begin{array}{cc}
\thickone -p_M a_M a^\dag_M & 
 (1-p_M) a_M^\dag \\
p_M a_M & p_M \thickone
\end{array}
\right)
\right]
\label{TF3}
\end{equation}
To obtain (\ref{TF3}) from (\ref{TF2}) it may be checked that for
arbitrary commuting
 operators $x_i$ and $y_i$ and arbitrary operators $z_i$, one has the identity
\begin{eqnarray}
\begin{array}{c}
 ( x_1,\ x_2) \\
\mbox{ }
\end{array}
\left(
\begin{array}{cc}
y_1 & y_2 \\
y_3 & y_4
\end{array}
\right)
\left( \begin{array}{c}
	z_1\\
	z_2
	\end{array} \right) 
= \mbox{Trace} \left[
\left(
\begin{array}{cc}
y_1 & y_2 \\
y_3 & y_4 
\end{array}
\right)
\left(
\begin{array}{cc}
x_1 z_1 & x_2 z_1 \\
x_1 z_2 & x_2 z_2 
\end{array}
\right)
\right]\;.
\end{eqnarray}

We stress here that (\ref{TF3}) is merely a convenient way of writing
the sums of products of the operators $\thickone, a_\mu, a_\mu^\dag$
which form the transfer matrix. The two by two matrices appearing in
(\ref{TF3}), whose elements are made up of the operators $\thickone,
a_\mu, a_\mu^\dag$ may be thought of as acting in some auxiliary
space.  The trace is carried out in this auxiliary space and not in
the space in which the operators $\thickone, a_\mu, a_\mu^\dag$ act.

In a similar fashion one can construct the transfer matrix for
backward updating, this time using (\ref{hmu2}) in (\ref{TB1}). One
obtains
\begin{equation}
\hat T_B = \mbox{Trace} \left[
\left(
\begin{array}{cc}
\thickone -p_M a_M a^\dag_M & 
 p_M a_M \\
a_M^\dag -p_M a_M a_M^\dag a_M^\dag & p_M a_M a_M^\dag
\end{array}
\right)
\prod_{\mu=1}^{M-1}
\left(
\begin{array}{cc}
\thickone -p_{M-\mu} a_{M-\mu} a^\dag_{M-\mu} & 
p_{M-\mu} a_{M-\mu}\\
(1-p_{M-\mu}) a_{M-\mu}^\dag & p_{M-\mu} \thickone
\end{array}
\right)
\right]
\label{TB3}
\end{equation}

Finally, to construct the transfer matrix for parallel dynamics $\hat T_P$ one
should first realise that it is closely related to $\hat T_F$ since
for forward updating each particle is unaffected by the results of
previous updates in the sequence, except for the final particle
$M$. After a little reflection it can be confirmed that
\begin{equation}
\hat T_P = \mbox{Trace} \left[
\prod_{\mu=1}^{M}
\left(
\begin{array}{cc}
\thickone -p_\mu a_\mu a^\dag_\mu & 
 a_\mu^\dag-p_\mu a_\mu a^\dag_\mu a^\dag_\mu \\
p_\mu a_\mu & p_\mu a_\mu a_\mu^\dag
\end{array}
\right)
\right]
\label{TP3}
\end{equation}

\subsection{Proof of the Steady State}
It was stated in section 2.1 that the steady state is
given by (\ref{Ffac}), with $f_\mu(n_\mu)$ obeying
(\ref{fP}) for parallel dynamics, (\ref{fF}) for forward sequential dynamics and
(\ref{fB}) for backward sequential dynamics. The proof is quite similar in all three
cases. We shall describe it in detail first for the case of parallel
dynamics where the transfer matrix has the form (\ref{TP3}). At the
end of the subsection we shall return to forward sequential
then backward sequential dynamics.

We first note that the vector space spanned by $| n_1,\ldots, n_\mu
\rangle $ is a tensor product of $M$ spaces each with basis vectors $|
n_\mu >$ where $n_\mu = 0 \cdots \infty$
\begin{equation}
| n_1,\ldots, n_\mu \rangle 
= | n_1 > \otimes | n_2 >  \cdots \otimes | n_M > \;.
\label{tprod}
\end{equation}
Thus  the action of the transfer matrix for
parallel dynamics (\ref{TP3}) on (\ref{tprod}) may be written as
\begin{equation}
\hat T_P | n_1, \ldots , n_M \rangle = \mbox{Trace} \left[
\prod_{\mu=1}^{M}
\left(
\begin{array}{cc}
\left(\thickone -p_\mu a_\mu a^\dag_\mu \right) | n_\mu \rangle  & 
( a_\mu^\dag-p_\mu a_\mu a^\dag_\mu a^\dag_\mu )| n_\mu \rangle\\
p_\mu a_\mu | n_\mu \rangle & p_\mu a_\mu a_\mu^\dag | n_\mu \rangle
\end{array}
\right)
\right]
\label{TP4}
\end{equation}
where the product, in fact, indicates a tensor product in the
configuration space (as well as a usual product in the auxiliary space
of two by two matrices) and the operators $\thickone, a_\mu,
a_\mu^\dag$  act on the space spanned by $| n_\mu \rangle$. To keep
the notation light we have not explicitly indicated these matters, but
the meaning is clear.

We are now in a position to prove the expression for the steady state
given by (\ref{fP}) and (\ref{Ffac}).  If the steady state is of the
factorised form $(\ref{Ffac})$ we have
\begin{equation}
\langle F | = \sum_{ \{ n_\mu \} }
\left[\ f_1(n_1) \langle n_1 | \ \right] \otimes
\left[\ f_2(n_2) \langle n_2 | \ \right]   \cdots \otimes 
\left[\ f_M(n_M) \langle n_M | \ \right]
\label{Ffacvec}
\end{equation}
and one finds from (\ref{TP4},\ref{Ffacvec}) that
\begin{equation}
\langle F |\hat T_P | n_1, \ldots , n_M \rangle = 
\mbox{Trace} \left[
\prod_{\mu=1}^{M}
B_\mu( n_\mu )
\right]\; ,
\label{TraceB}
\end{equation}
where
\begin{eqnarray}
B_\mu(0) &=&
\left(
\begin{array}{cc}
f_\mu(0) & 0 \\
p_\mu f_\mu(1) & 0 \\
\end{array}
\right) \label{Bdef1}\\
B_\mu(1) &=&
\left(
\begin{array}{cc}
(1-p_\mu)f_\mu(1) & f_\mu(0) \\
p_\mu f_\mu(2) & p_\mu f_\mu(1) \\
\end{array}
\right) \label{Bdef2} \\
\mbox{and for}\;\;\;n >1\;\;\; B_\mu(n) &=&
\left(
\begin{array}{cc}
(1-p_\mu)f_\mu(n) & (1-p_\mu) f_\mu(n-1) \\
p_\mu f_\mu(n+1) & p_\mu f_\mu(n) \\
\end{array}
\right)\;.
\label{Bdef3}
\end{eqnarray}
Inserting the expressions for $f_\mu(n)$ given in (\ref{fP}),
and employing the Heaviside function we 
may rewrite (\ref{Bdef1}--\ref{Bdef3}) as
\begin{equation}
B_\mu(n) =f_\mu (n) A_\mu(n)
\label{BfA}
\end{equation}
where
\begin{equation}
A_\mu (n) = \left(
\begin{array}{cc}
1-p_\mu\theta(n) & p_\mu\theta(n) \\
1-p_\mu\theta(n) & p_\mu\theta(n)
\end{array}
\right)\; .
\label{Amu}
\end{equation}
To   prove the steady state (\ref{Ffac}) we must show
 that $\langle F |\hat T_P | n_1, \ldots , n_M \rangle = 
\prod_{\mu=1}^{M} f_{\mu} ( n_{\mu} )$.
Thus,  on inserting (\ref{BfA}) into (\ref{TraceB}), we see that
it remains to show
\begin{equation}
\mbox{Trace} \left[
\prod_{\mu=1}^{M}
A_\mu( n_\mu )
\right] = 1 
\label{TraceA}\;.
\end{equation}

In order to do this we seek a similarity transformation
\begin{equation}
\tilde A_\mu(n) = L_{\mu} A_\mu(n) R_{\mu+1}
\;\;\;\mbox{with}\;\;\; R_\mu L_\mu = \thickone
\label{simtran}
\end{equation}
that puts $A_\mu(n)$ into the form
\begin{equation}
\tilde A_\mu(n) =
\left(
\begin{array}{cc}
1 & x \\
0 & 0
\end{array}
\right)
\label{require}
\end{equation}
where $x$ is any number. The trace of a product of such matrices
(\ref{require}) is clearly
unity thus satisfying (\ref{TraceA}). A similarity transformation which fulfills
(\ref{simtran}, \ref{require}) is straightforward
 to construct. Taking
\begin{equation}
R_\mu = \left(
\begin{array}{cc}
1 & p_\mu \\
1 & -(1-p_\mu)
\end{array}
\right)
\;\;\;\mbox{and}\;\;\; 
L_\mu  = \left(
\begin{array}{cc}
1-p_\mu & p_\mu \\
1 & -1
\end{array}
\right)\; ,
\label{simmat}
\end{equation}
it is easy to verify that $R_\mu L_\mu= \thickone$ and
\begin{eqnarray}
\tilde A_\mu(n)& =& L_{\mu} A_\mu(n) R_{\mu+1 }=
 \left(
\begin{array}{cc}
1 & p_{\mu+1}-p_\mu \theta(n)  \\
0 & 0
\end{array}
\right) 
\end{eqnarray}
hence (\ref{fP}) is proven.

In a similar manner the steady state for forward updating may be
proven. Here we only provide a few key points. Following the steps
(\ref{TP4}) through to (\ref{Amu}), now using (\ref{TF3}) as the
transfer matrix one finds that (\ref{Amu}) holds for $\mu < M$ and
\begin{equation}
A_M (n) = \left(
\begin{array}{cc}
1-p_M\theta(n) & p_M\theta(n) \\
1-p_M & p_M
\end{array}
\right)\; .
\label{AMF}
\end{equation}
The similarity transformation (\ref{simtran}) then gives
\begin{equation}
\tilde A_M (n) = \left(
\begin{array}{cc}
1 &\;\;\; p_1 -p_M^2 -p_M(1-p_M)\theta(n) \\
0 &\;\;\; p_M(1-\theta(n))
\end{array}
\right)
\label{AMFtilde}
\end{equation}
whilst the other
 $\tilde A_\mu (n)$ have the form
(\ref{require}). The trace of a product of these $\tilde A_\mu (n)$
again gives unity.

For backward updating, using (\ref{TB3}) for the transfer matrix, one
obtains for $\mu < M$
\begin{equation}
A_\mu (n) = \left(
\begin{array}{cc}
1-p_\mu\theta(n) &1-p_\mu \\
p_\mu\theta(n) & p_\mu
\end{array}
\right)
\label{AmuB}
\end{equation}
and for $\mu=M$
\begin{equation}
A_M (n) = \left(
\begin{array}{cc}
1-p_M\theta(n) &1-p_M\theta(n) \\
p_M\theta(n) & p_M\theta(n)
\end{array}
\right)
\label{AMB}
\end{equation}
A suitable similarity transform is now
\begin{equation}
\tilde A_\mu(n_\mu) = R_{\mu+1}^{\rm T} A_\mu(n_\mu) L_{\mu}^{\rm T}
\label{simtranT}
\end{equation}
where $R_{\mu},  L_{\mu}$ are still given by (\ref{simmat}). One then obtains
for $\mu< M$
\begin{equation}
\tilde A_\mu (n) = \left(
\begin{array}{cc}
1 &\;\;\; 0 \\
p_{\mu+1} -p_{\mu}^2 -p_{\mu}(1-p_{\mu})\theta(n) &\;\;\; p_\mu(1-\theta(n))
\end{array}
\right)
\label{AMBtilde}
\end{equation}
and
\begin{equation}
\tilde A_M (n) = \left(
\begin{array}{cc}
1 & 0 \\
p_{1} -p_{M}\theta(n) &0
\end{array}
\right)\;.
\label{AMB2tilde}
\end{equation}
It is easy to convince oneself that the trace of a 
product of matrices of form (\ref{AMBtilde})
with a single matrix of form (\ref{AMB2tilde}) yields unity,
thus completing the proof of the steady state for
backward updating.
\nsection{Generalisation to hopping probabilities dependent on empty
space in front of particles} 
In this section we consider hopping probabilities that depend on the
number of empty sites in front of a particle. For this dynamics with
random sequential updating, it is known  that the
steady state is  given by a product measure \cite{Spitzer}. 
Here for the disordered
case we restrict our attention to parallel updating although it is straightforward to generalise to
ordered sequential updating.  Thus for parallel updating, at each
time-step particle $\mu$ will hop forward with probability
$p_\mu(n_\mu)$, where we recall $n_\mu$ is the number of empty sites
in front of particle $\mu$ so that clearly $p_\mu(0)=0$.  In a traffic
model suitably chosen hopping probabilities could mimic the effect of
deceleration when another car is encountered.

It turns out that the steady state again has the form
(\ref{Ffac}).  It will be shown in this section that
\begin{eqnarray}
f_\mu( n )
 &=& (1- p_\mu(1))\;\;\;\mbox{for}\;\;\; n =0 \nonumber \\
 &=&  \frac{1-p_\mu(1)}{1-p_\mu(n)}
\prod_{k=1}^{n}\frac{1-p_\mu(k)}{p_\mu(k)}
\;\;\;\mbox{for}\;\;\; n >0 \;.
\label{fPgen}
\end{eqnarray}
In order to prove (\ref{fPgen}) we generalise the operators used in
(\ref{hmu}--\ref{lower2}) by defining new operators
$a_{\mu}(n)$,$b_{\mu}(n)$:
\begin{eqnarray}
a_\mu(n)  | n_{1}, \ldots, n_\mu, \ldots, n_{M} >
&=& | n_{1}, \ldots, n_\mu +1, \ldots, n_{M} > 
 \;\;\;\mbox{if}\;\;\;
n_\mu = n   \label{raisegen}\\
&=& 0 \;\;\;\mbox{otherwise}\;\;\;\\
b_{\mu}(n) | n_{1}, \ldots, n_\mu, \ldots, n_{M} >
&=& | n_{1}, \ldots, n_\mu, \ldots, n_{M} >
 \;\;\;\mbox{if}\;\;\;
n_\mu = n   \label{lowergen}\\
&=& 0 \;\;\;\mbox{otherwise}\;.
  \label{lowergen2}
\end{eqnarray}
The corresponding operator to (\ref{hmu}) is now given by
\begin{equation}
h_{\mu-1 \ \mu} = \thickone - \sum_{n=1}^{\infty}
 p_{\mu}(n)\ b_{\mu}(n) +  a_{\mu-1}^{\dag}
\sum_{n=1}^{\infty} p_{\mu}(n+1)\ a_\mu(n)\;,
\label{hmugen}
\end{equation}
where $a_{\mu-1}^{\dag}$ is still defined by
(\ref{lower},\ref{lower2}).  Using the same procedure outlined in
section 4 the transfer matrix (for parallel dynamics) may be
constructed as
\begin{equation}
\hat T_P = \mbox{Trace} \left[
\prod_{\mu=1}^{M}
\left(
\begin{array}{cc}
\thickone -\sum_{n=1}^\infty p_\mu(n) b_\mu(n) & 
 a_\mu^\dag-\left[ \sum_{n=1}^{\infty} p_\mu(n) b_\mu(n)\right] \ a^\dag_\mu \\
\\
\sum_{n=1}^\infty p_\mu(n+1)\ a_\mu(n) &
\left[ \sum_{n=1}^\infty p_\mu(n+1) a_\mu(n)\right] a_\mu^\dag
\end{array}
\right)
\right]\;.
\label{TPgen}
\end{equation}
Using this transfer matrix and assuming that the steady state is of
the form (\ref{Ffac}), in analogy with section 4 one arrives at (\ref{TraceB}) where now
\begin{equation}
B_\mu(n) =
\left(
\begin{array}{cc}
f_\mu(n)(1-p_\mu(n)) &\;\;\; f_\mu(n-1)(1-p_\mu(n-1))\theta(n) \\
f_\mu(n+1) p_\mu(n+1)  &\;\;\; f_\mu(n) p_\mu(n)  \\
\end{array}
\right)\;.
\end{equation}
and it should be  borne in mind that $p_\mu(0)=0$.
Inserting (\ref{fPgen}) one finds  that it remains to
show condition (\ref{TraceA}) is satisfied where this time
\begin{equation}
A_\mu (n) = \left(
\begin{array}{cc}
1-p_\mu(n) & p_\mu(n) \\
1-p_\mu(n) & p_\mu(n)
\end{array}
\right)
\end{equation}
It is easy to check that inserting $R_\mu(n_\mu) L_\mu(n_\mu)$
($=\thickone$) before each $A_\mu(n_\mu)$ in the product
(\ref{TraceA}), where
\begin{equation}
R_\mu(n_\mu) = \left(
\begin{array}{cc}
1 & p_\mu(n_\mu) \\
1 & -(1-p_\mu(n_\mu))
\end{array}
\right)
\;\;\;\mbox{and}\;\;\; 
L_\mu(n)  = \left(
\begin{array}{cc}
1-p_\mu(n_\mu) & p_\mu(n_\mu) \\
1 & -1
\end{array}
\right)\; ,
\label{simmatgen}
\end{equation}
yields the required similarity transformation.

\nsection{Conclusion} 
In this paper it has been shown that analytical results may be
obtained for the asymmetric exclusion model with parallel dynamics
which forms the basis for many discrete models of traffic
flow. Ordered sequential dynamics
have also been treated using similar techniques.
Previously, the steady state for the situation where at each
parallel time-step each particle attempts forward with probability $p$
had been solved \cite{SSNI}. Here we have generalised that model to
one where at each time-step each particle $\mu$ attempts a hop forward
with probability $p_\mu (n_\mu)$, an arbitrary
function of $n_\mu$ the number of empty sites immediately ahead of
particle $\mu$. The method used to prove the steady state
(\ref{fPgen}) was to construct the transfer matrix using a technique
inspired by \cite{HN}.

A detailed analysis was made of the case where $p_\mu$ did not depend
on $n_\mu$, and it was shown that a phase transition analogous to Bose
condensation occurs, as it did for a random sequential updating
\cite{MRE}. In particular it is interesting to note equation
(\ref{vel3}) which shows the velocity in the particle system is
intimately related to the fugacity (or equivalently chemical
potential) when the system is viewed as a Bose system.  Since in (single
component) equilibrium systems the thermodynamic phase is determined
by minimising the chemical potential, the suggestion is that
in simple traffic  flow models a principle of minimisation of
velocity pertains.

A simple illustration of this principle is the example of a single
slow particle with hopping rate $c$ while all the other particles
have hopping rate $p$ \cite{MRE}. Then in a phase where there is a tailback
behind the slow particle and empty space in front, the velocity is $c$
whereas in a congested phase the velocity should be given by 
$(\ 1-\sqrt{1-4p \rho(1-\rho)}\ )/(2 \rho)$ \cite{SSNI}.
Choosing the phase with the minimum velocity
 yields the critical density $\rho_c = (p-c)/(p-c^2)$.
This expression is, in fact, exact as can be checked by using the results
of section 3. It is important to ascertain whether
a principle of minimisation of
velocity holds in more complicated traffic flow models.

It was also interesting to note that parallel dynamics and forward
sequential dynamics have the same thermodynamic behaviour e.g. the
same critical density.  One expects this since a time-step of forward
updating and a parallel time-step only differ in last update of the
sequence in forward dynamics.  However backward updating has a
distinct critical density as does random sequential updating.  Comparing
the three critical densities reveals that the highest is for
backward dynamics, the lowest is for random sequential dynamics and
 parallel dynamics lies in between. This confirms that
backward dynamics is the most efficient updating scheme in terms of
throughput and random sequential is the worst.

It would be interesting to explore other realisations for the case of
general $p_\mu(n_\mu)$ considered in section 5.  In particular it
should be possible to analyse the effect of a braking distance for
each particle.

The present work narrows the gap between {\it bona fide} models of
traffic flow, such as that of Nagel and Schreckenberg  or
more sophisticated models \cite{Nagel}, and the simple particle hopping models for
which exact results are possible.  In order to close the gap
further, analytical results are desirable for models where particles
can hop more than one lattice site at each update.  Some progress has
already been made towards this \cite{BPSS}.

It would also be of interest to analyse more complicated properties of
the system such as relaxation to the steady state, as studied
numerically in \cite{KF,KCW}, or else the diffusion constants of
particles \cite{DEM,DEMall}.

\nsubsecnn{Acknowledgments}
I thank F. H. L. Essler, H. Hinrichsen and G. M. Sch\"utz
for useful discussions and especially
V. Hakim for patient explanation of the technique used in \cite{HN}.

\appendix{A: Evaluation of the Integral (\ref{int})}
\setcounter{equation}{0}
\def\theequation{A\arabic{equation}}
In this appendix we derive the expansion
(\ref{gamm<1}--\ref{gamm>1}) of the  integral (\ref{int})
given by
\begin{equation}
I(v)=\frac{\gamma+1}{(1-c)^{\gamma+1}}
\int_{c}^{1} dp \  \frac{v(p-c)^\gamma}{p-v}\; .
\label{Intdef}
\end{equation}
By first defining
\begin{equation}
v = c \exp( -\epsilon)\;,
\end{equation}
(\ref{Intdef}) becomes
\begin{equation}
I(\epsilon)=\frac{\gamma+1}{(1-c)^{\gamma+1}}
\int_{0}^{1-c} dp \  
\frac{c \exp(-\epsilon)p^\gamma}{p+c-c \exp(-\epsilon)}
\label{Intepsdef}
\end{equation}
and we may follow an analysis similar to that of Robinson \cite{Rob}.
We  make the Mellin transformation
\begin{equation}
{\cal I}(s) = \int_{0}^{\infty} d\epsilon \ I(\epsilon) \epsilon^{s-1}\; ,
\label{Mellindef} 
\end{equation}
which may be expanded to yield
\begin{equation}
{\cal I}(s)=
\frac{\gamma+1}{(1-c)^{\gamma+1}}
\int_{0}^{1-c} dp \ p^\gamma
\sum_{n=0}^{\infty} \left( \frac{c}{p+c} \right)^{n+1}
\ \frac{\Gamma(s)}{(n+1)^s} \;,
\label{Int1}
\end{equation}
where $\Gamma(s)$ is the usual Gamma function defined
by
\begin{equation}
\Gamma(s) = \int_{0}^{\infty}d \epsilon\; \exp( -\epsilon)\ \epsilon^{s-1}\;.
\end{equation}
The inverse transformation
is
\begin{equation}
 I(\epsilon) = \frac{1}{2\pi i}\int_{x-i \infty}^{x+i \infty} d s
\ {\cal I}(s) \epsilon^{-s}\; ,
\label{Mellininv} 
\end{equation}
where $x$ is a real constant chosen so that the contour of integration
is to the right of any pole. The integral can be evaluated by closing
the contour and using the calculus of residues.  The analytic
structure of
\begin{equation}
{\cal I}(s) = g_\gamma(s) \Gamma(s)\; ,
\end{equation}
where
\begin{equation}
g_\gamma(s) = \frac{\gamma+1}{(1-c)^{\gamma+1}}
\int_{0}^{1-c} dp \ p^\gamma
\sum_{n=0}^{\infty} \left( \frac{c}{p+c} \right)^{n+1}
\ \frac{1}{(n+1)^s}\; ,
\label{gdef}
\end{equation}
is as follows: $\Gamma(s)$ has simple poles at $s=-n$
where $n=0,1,\ldots\infty$
with residues $(-1)^n/n!$ and
$g_\gamma(s)$ has a simple pole at $s=-\gamma$.
To evaluate the residue of $g_\gamma(s)$ at $s=-\gamma$ we note the small $p$
behaviour of the sum involved in (\ref{gdef})
\begin{eqnarray}
\sum_{n=0}^{\infty} \left( \frac{c}{p+c} \right)^{n+1}
\ \frac{1}{(n+1)^s}
\end{eqnarray}
Thus, one obtains
\begin{eqnarray}
\lim_{s\to -\gamma} \left[ g_\gamma(s)\ (s+\gamma) \right]
&=& (1+\gamma) \left( \frac{c}{1-c}\right)^{1+\gamma} \Gamma(1+\gamma)\; .
\end{eqnarray}
and
\begin{equation}
 I(\epsilon)=
 (1+\gamma) \left( \frac{c}{1-c}\right)^{1+\gamma}
\Gamma(1+\gamma) \Gamma(-\gamma) \epsilon^\gamma
+ \sum_{n=0}^{\infty} \frac{g_\gamma(-n)}{n!} \ (-\epsilon)^n\;.
\label{Mellinres}
\end{equation}
The expansion (\ref{gamm<1},\ref{gamm>1}) is the first two terms in
(\ref{Mellinres}) where it can be computed that
\begin{eqnarray}
g_\gamma(0) &=& \frac{1+\gamma}{\gamma}\ \frac{c}{1-c} 
\label{g(0)} \\
g_\gamma(-1) &=& \frac{1+\gamma}{\gamma}\ \frac{c}{1-c}\left[ 1 +  
\frac{\gamma}{\gamma-1}\ \frac{c}{1-c}\right]\;.
\label{g(-1)}
\end{eqnarray}
For $\gamma \to 1$, the singularity in (\ref{g(-1)}) cancels with the
singularity in the first term in the rhs of (\ref{Mellinres}) and one
obtains
\begin{eqnarray}
I(\epsilon) &=& g_\gamma(0)+ \lim_{\gamma \to 1}
\left[  
 (1+\gamma) \left( \frac{c}{1-c}\right)^{1+\gamma}
\Gamma(1+\gamma) \Gamma(-\gamma) \epsilon^\gamma
-g_\gamma(-1)\epsilon  \right] + \dots
\nonumber \\
&=&
g_\gamma(0) + 2 \left(\frac{c}{1-c}\right)^2 \epsilon \ln \epsilon
+ {\cal O}(\epsilon)\; .
\end{eqnarray}
%

\end{document}